# Evidence for Torque Caused by a Magnetic Impulse on a Nonmagnetic Torsion Pendulum

Jean Paul Mbelek

*Sangha Center for Astronomy, Astrophysics and Cosmology, Sangha, Mali*

Abstract : I have performed an experiment which is a variant of the one suggested recently by F. O. Minotti and T. E. Raptis (arXiv: 1310.5029). The aim of this experiment is to check the generation of a pulsed gravitational potential, $\chi$, by a transient magnetic field as predicted by a scalar tensor theory (STT) of gravity. Such a STT allows an enhanced coupling of its fundamental long-range real scalar field, $\phi$, as well as its external scalar field, $\psi$, to the electromagnetic (EM) field, according to the author's previous papers. Several possible sources that could mimic the predicted effect have been ruled out. Finally, it seems likely that the predicted effect is there with the right order of magnitude, as expected.

I. Introduction

Sometime ago, along with M. Lachièze-Rey, we have suggested to add an external scalar field, $\psi$, to the 5D Kaluza-Klein (KK) theory as a possible solution to cure the instability of its action after dimensional reduction [1]. By doing so, it became possible through the source term of the $\psi$ field to couple both the $\psi$ field and the internal KK scalar field, $\phi$, to matter as well as to the EM field through coupling functions depending on both scalar fields and the temperature. Moreover, solutions involving enhanced gravitational effects as compared to general relativity (GR) become possible within such a framework. The KK$\psi$ theory, as this particular STT was initially dubbed, was successfully applied to the outstanding problem of the discrepant laboratory G measurements [2,3] as well as to the cosmological variation of the fine structure constant [4] and the modelling of the rotational curves of spiral galaxies [5].

In the weak field and low velocity limit, some of the coupling constants have been determined in the first-order approximation from fits to the data. In particular, the coupling of the $\phi$ field to the EM field can be translated in terms of a universal constant of the dimension of a force, $\mathcal{F}$. This constant, $\mathcal{F} \approx [(5.44\pm0.66)\times10^{-6} \text{fm}\cdot\text{TeV}^{-1}]^{-1} \approx (2.99\pm0.36)\times10^{13}\text{N}$, has been estimated from the fit of the measured gravitational constant, G, as a function of the geomagnetic potential at the laboratory location [2,3]. Recently, F. O. Minotti argued that the KK$\psi$ theory could account for the experimental results [6,7] concerning the appearance of unusual forces on asymmetric electromagnetic resonant cavities [8,9]. On the theoretical grounds, F.O Minotti [8] has shown that an additional scalar field minimally coupled to gravity may help to reconcile a Brans-Dicke coupling constant, $\omega$, of the order unity with the solar system bounds (see also [10]).

Furthermore, on the experimental grounds, F. O. Minotti and T. E. Raptis suggested two different experiments that could test the KKψ theory in the laboratory. The first one deals with the possible variation of the amplitude of a laser beam propagating within an optical fiber exposed to a static electric or magnetic field [11], and the other one consists in exciting a pendulum oscillation by turning on and off the current in a coil [12]. Hereafter, we will focus our attention on the latter proposal which implies the swing right to left of the pendulum. However, it turns out that it is difficult to separate the swing right to left or back and forth of the pendulum from vibrations and many other extraneous disturbances. Instead, the torsion motion of the pendulum is well determined and may be easily separated from vibrations and several extraneous disturbances. Indeed, I have been able to perform the latter experiment accurately by using materials and instruments that are available in many academic laboratories. It turns out that the theoretical predictions of the KKψ theory are in good agreement with the experimental data.

In what follows, the experimental setup is described in Section II and then a theoretical interpretation is provided in the framework of the KKψ theory in Section III. A critical discussion of the experiment is carried out in Section IV. Finally, the conclusion is given in Section V.

II. Experimental setup

The system of interest consists of a pendulum whose bob is a concave mirror, $M_1$, of mass m = 9g and radius R = 3.5cm, attached to a wire of length $l$ = OM = 25cm, where M denotes the center of mass of $M_1$. The pendulum is enclosed in a closed chamber provided with two fine openings $A_1$ and $A_2$ such that $(A_1MA_2) = \pi/2$ rad (see Fig. 2). A few meters further, there is a power supply which constitutes an electric circuit with a switch and a solenoid (see Fig. 1 and 2); the solenoid is closer to the chamber of the pendulum. The power supply of emf V fixed to 30 Volts (adjustable 0-30V/0-5A, a direct current voltage source with a residual ripple below a maximum of 1 mV rms, the current ripple less than 3 mA) and the internal resistance $R_E$ = 1.8Ω drives a current, i = i(t), through the solenoid of inductance L = 0.8H and resistance $R_L$ = 10Ω. The concave mirror, $M_1$, is absolutely free from magnetic material. This can be seen by putting a strong permanent magnet (neodymium-iron-boron) on the mirror and noting that no force is exerted on the latter. The same test is applied to the torsion wire, the adjust knob and the chamber as well with the same null result.

The solenoid, the chamber containing the pendulum and the DC power supply, each of these elements stands on its own vibration-free platform. As a consequence, vibrations are strongly damped, so that the pendulum behaves mostly like a torsion pendulum. The angular deviation θ of the pendulum implies a linear deflection on the screen equal to d = 2θ×(MM′ + M′M″); MM′ = 3.6m and M′M″ = 5m. Besides, the position of the solenoid can be adjusted with respect to the bob of the pendulum in order to maximize the amplitude of the deviation, d, of the light spot on the screen.

Three kinds of motions are possible, although only two of them can be observed through the motion of the light spot of the reflected light beam on the screen. Let us point out that the swing from left to right of the pendulum, whose axis of rotation is the x axis and which is described by the angle α, cannot be displayed on the screen. The two other motions, that we are interested in and that can be seen on the screen, are the swing back and forth of the pendulum whose axis of rotation is the y axis and which is described by the angle β and the torsion of the pendulum whose axis of rotation is the z axis and which is described by the angle θ (see Fig. 1).

Under the influence of the magnetic impulse generated by the coil, the nonmagnetic pendulum starts to execute damped torsion oscillations at a fixed frequency, $\nu_0 \approx 0.14$Hz, and to a lesser extent to oscillate about the equilibrium position, swinging back and forth at a fixed frequency close to 1Hz. By changing solely the orientation of the solenoid, no noticeable modification of the deflection d is observed on the screen. The different coil orientations yield the same result for any given distance between the coil and the bob, in consistency with the scalar nature of the fields underlying the observed effect. However, by taking the measurements, we ensured that the axis of the solenoid be aligned as closely as possible with the normal to the bob. Besides, the bob axis is always perpendicular to the z axis. When the magnetic core is pulled into the solenoid coil, one can see a significant deviation of the light spot on the screen. Conversely, when the magnetic core is pulled out of the solenoid coil, no deviation of the light spot is observed at all on the screen.

The magnetic core of the solenoid coil is made of mu-metal ($\mu_r \sim 10^3$). The inductance of the solenoid coil has been measured with the help of a digital multimeter. Hence, the relative permeability of the core of the coil can be estimated in the first-order approximation from the relation $\mu_r \approx L(\ell_{coil}/\mu_0 N^2 \pi R_{core}^2)[1 - (8R_{core}/3\pi \ell_{coil})]$. One finds $\mu_r \approx 6.7 \times 10^3$, where L = 0.8H, N = 200, $R_{core}$ = 1.3cm and $\ell_{coil}$ = 22.5cm. As expected, the value found for $\mu_r$ falls well within the range allowed for mu-metal as can be found in the literature.

III. Theoretical interpretation in the framework of STT

The coupling of the magnetic field **H** to the scalar field φ yields in the weak field and low velocity limit a gravitational potential χ whose equation reads in vacuum or air [4-6]

$$\Box \chi = -(2\pi\mu_0/\mathcal{F})\mathbf{H}^2. \quad (1)$$

Within a magnetized medium of relative permeability $\mu_r$ we propose to generalize the above equation as follows:

$$\Box \chi = -(2\pi\mu_0/\mathcal{F}\mu_r)\mathbf{H}^2. \quad (2)$$

For our purpose, we should also include the geomagnetic induction, $\mathbf{B}_\oplus$, so that Eq. (2) becomes

$$\Box \chi = -(2\pi\mu_0/\mathcal{F}\mu_r)[\mathbf{H} + (\mathbf{B}_\oplus/\mu)]^2, \quad (3)$$

where **B** and **H** = **B**/μ denote the magnetic induction and the magnetic field of the solenoid coil, respectively; $\mu_0 = 4\pi \times 10^{-7}$SI and $\mu = \mu_r\mu_0$ are the permeabilities of the vacuum and the magnetic core of the inductor coil, respectively; $\mu_r$ is the relative permeability of the magnetic core.

When the magnetic core is placed off the solenoid, then the strength of the magnetic induction drops to $B = \mu_0 H$ which may be less in magnitude than $B_\oplus$ so that Eq. (1) reduces to

$$\Box\chi = -(2\pi/\mathcal{F})B_\oplus^2/\mu_0. \quad (4)$$

This implies that the pendulum will remains at its equilibrium position, in accordance with the experiment. Alternatively, when the magnetic core is placed inside the solenoid, then the strength of the magnetic induction becomes much larger than in the previous case by a factor $\mu_r \sim 5\times10^3$, that is, $B = \mu_r\mu_0 H$, which is by far greater in magnitude than $B_\oplus$, so that Eq. (1) reduces to

$$\Box\chi = -(2\pi\mu_0/\mathcal{F}\mu_r)(B/\mu)^2. \quad (5)$$

The above equation yields the following solution in terms of the retarded potential:

$$\chi(M,t) = 1 - (\mu_0/2\mathcal{F}\mu_r)\iiint_{(coil)} [(B(t-PM/c)/\mu)^2/PM]d^3\mathbf{OP} \approx 1 - \tfrac{1}{2}(L/\mu_r^2)i^2_{(t-r/c)}/\mathcal{F}r, \quad (6)$$

where $\iiint_{(coil)} B^2/2\mu\, d^3\mathbf{r} = \tfrac{1}{2} L i^2$ denotes the magnetic energy stored in the volume of the solenoid of inductance L through which a current i flows; $r = SM = 23\,cm$, where S denotes the center of mass of the solenoid. Thus the total conservative force $\mathbf{F}'$ involved both by the gravitational potential $\chi$ and the external scalar field $\psi$ that can be exerted by the solenoid on the bob of mass m reads

$$\mathbf{F}' = -mc^2\nabla(\chi + k\psi) \approx -mc^2\nabla\chi \approx -\tfrac{1}{2}\,mc^2(L/\mu_r^2)i^2_{(t-r/c)}\mathbf{u_r}/\mathcal{F}r^2, \quad (7)$$

where $k \approx 0.05$ [3].

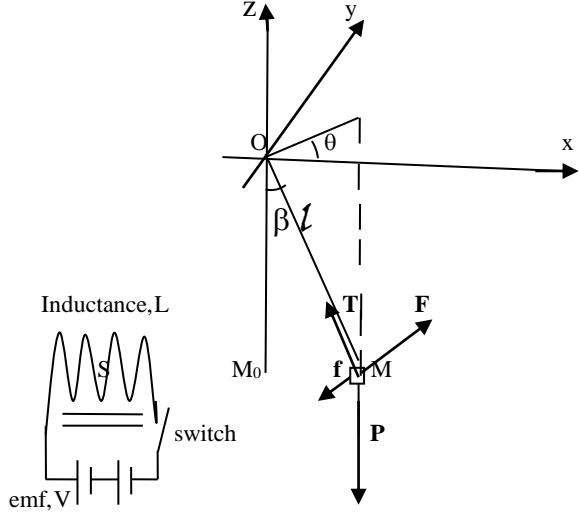

**Fig. 1.** S(X, Y, Z) denotes the center of mass of the solenoid, M(x(t), y(t), z(t)) the center of mass of the bob of the pendulum and $M_0(0, 0, -l)$ the static equilibrium position of the bob of the pendulum in the coordinates system (Ox, Oy, Oz); $Z \approx -l$ and $X \approx -r$. The z axis is vertical whereas the x and y axes define the horizontal plane. The points $M_0$ and M denote the equilibrium position and the position of the center of mass of the bob at time t, respectively. The bob of the pendulum is subject to its weight, **P** = m**g**, the tension of the wire, **T**, the friction force, **f**, and the force terms due to the scalar fields, $\mathbf{F} = -mc^2\nabla(\chi + k\psi) + km\mathbf{v}d\psi/dt$. When the switch is on the charging current reads $i = I(1 - e^{-t/\tau})$, where $I = V/(R_E + R_L)$ and $\tau = L/(R_E + R_L)$. When the switch is off, the discharging current reads $i = I e^{-(t-t_0)/\tau}$.

Besides, the external scalar field $\psi$ also involves a drag force term $\mathbf{F}'' = km(d\psi/dt)\mathbf{v}$ [5]. Hereafter we will model the bob as a thin disk of radius R, so that $J = ¼ mR^2$. Also, the pendulum swings back and forth at a frequency close to 1 Hz, which is just the value predicted by the formula of the simple gravity pendulum $\nu_S = (g/l)^{1/2}/2\pi$. Finally, the radius of the core of the coil being $R_{core} \approx 1$ cm, only a small fraction of the surface of the bob is permeated by the magnetic field. Based on these observations, we will consider equations where the bob is assimilated to a point mass as good approximation for our purpose. Above all, the correct procedure gives, in the first-order approximation, MP << SM, a similar torque to that of the point mass[1]. According to Newton second law, on account of the torsion torque

---

[1] Since MP ≤ (MP)$_{max}$ = R = 3.5 cm << r = SM = 23 cm, it follows **SP** ≈ **SM**. Hence, one obtains the following estimate of the torque:

$\iint_{(bob)} \mathbf{OP} \times (m/\pi R^2)c^2\nabla\chi \, dS = \iint_{(bob)} (\mathbf{OS} + \mathbf{SP}) \times (m/\pi R^2)c^2\nabla\chi \, dS$

$= (mc^2/\pi R^2)\iint_{(bob)} \mathbf{OS} \times \nabla\chi \, dS = ½ (L/\mu_r^2)I^2 (mc^2/\pi R^2 \mathcal{F}) \mathbf{OS} \times \int_0^R 2\pi \, MP \, (\mathbf{SP}/SP^3) dMP$

$\approx (L/\mu_r^2)I^2 (mc^2/R^2 \mathcal{F}) \mathbf{OS} \times \int_0^R MP \, (\mathbf{SM}/r^3) dMP = (L/\mu_r^2)I^2 (mc^2/R^2 r^3 \mathcal{F}) \mathbf{OS} \times \mathbf{SM} \int_0^R MP \, dMP$

$= ½ (L/\mu_r^2)I^2(mc^2/r^2\mathcal{F}) \mathbf{OS} \times \mathbf{u_r}$,

which is merely the point mass torque approximation used in Eq. (8).

and the torques due to the scalar field $\psi$, the potential $\chi$ and the air drag, the equation of motion of the pendulum reads

$$d\boldsymbol{\sigma}/dt = -h\mathcal{l}\boldsymbol{\Omega} - C\theta\mathbf{u}_z + \mathbf{OM}\times m\mathbf{g} - \iint_{(bob)} \mathbf{OP}\times mc^2\nabla(\chi + k\psi)\,dS/\pi R^2 + k(d\psi/dt)\boldsymbol{\sigma}$$

$$\approx -h\mathcal{l}\boldsymbol{\Omega} - C\theta\mathbf{u}_z + \mathbf{OM}\times m\mathbf{g} - \mathbf{OM}\times mc^2\nabla(\chi + k\psi) + k(d\psi/dt)\boldsymbol{\sigma}$$

(8)

where C, J and

$$\boldsymbol{\sigma} = J'(d\alpha/dt\,\mathbf{u}_x + d\beta/dt\,\mathbf{u}_y) + (J + J'\sin^2\beta)d\theta/dt\,\mathbf{u}_z$$

denote the torque constant of the wire, the moment of inertia with respect to the z axis and the angular momentum of the pendulum, respectively; $J' = m\mathcal{l}^2$. Also, $x = \mathcal{l}\sin\beta\cos\theta$, $y = \mathcal{l}\sin\beta\sin\theta$, $z = -\mathcal{l}\cos\beta$.

Besides, let us point out that the torsion motion is not an assumption in this study but merely an experimental fact. Indeed, in the absence of the magnetic field from the solenoid and by providing by hand the desired motion (either torsional or swinging motion) one can unambiguously identify clearly on the screen the kind of motion expected either for the torsion[2] or the swing back and forth of the pendulum[3]. Now, in this experiment, the swinging motion of the pendulum is not perceptible to the naked eye on the screen unlike the torsional motion. Thus, one may conclude that $\beta \ll 4\deg$ or otherwise stated $|x| \ll 1$ mm, so that $J'\sin^2\beta = m\mathcal{l}^2 x^2/Z^2 \approx mx^2 \ll J = \frac{1}{4}mR^2$. Since $|x| \ll \mathcal{l}$ and $t > 5\tau$, the projection of the above equation on the x, y and z axes yields

$$d^2\alpha/dt^2 + 2(\zeta - kd\psi/dt)d\alpha/dt + \omega_S^2\alpha \approx -(\tfrac{1}{2}\,Li^2_{(t-r/c)}/\mathcal{F}r\mu_r^2)[mc^2\mathcal{l}(Y + Z\beta\theta)/J'r^2], \quad (9)$$

$$d^2\beta/dt^2 + 2(\zeta - kd\psi/dt)d\beta/dt + [\omega_S^2 - (\tfrac{1}{2}\,Li^2_{(t-r/c)}/\mathcal{F}r\mu_r^2)(mc^2\mathcal{l}\,Z/J'r^2)]\beta$$

$$\approx (\tfrac{1}{2}\,Li^2_{(t-r/c)}/\mathcal{F}\mu_r^2 r)(mc^2\mathcal{l}X/J'r^2), \quad (10)$$

$$d^2\theta/dt^2 + 2(\lambda - kd\psi/dt)d\theta/dt + [\omega_T^2 - (\tfrac{1}{2}\,Li^2_{(t-r/c)}/\mathcal{F}|Z|\mu_r^2)(mc^2\mathcal{l}Xx/Jr^3)]\theta$$

$$\approx -(\tfrac{1}{2}\,Li^2_{(t-r/c)}/\mathcal{F}|Z|\mu_r^2)(mc^2\mathcal{l}\,Yx/Jr^3),$$

(11)

where, $\omega_S = (g/\mathcal{l})^{1/2}$ and $\omega_T = (C/J)^{1/2}$. Since the points S, $M_0$ and M are almost lined up,

---
[2] The spot of light moves left and right on the screen.
[3] The spot of light moves up and down on the screen.

$(\omega_S \ell/c)^2 \gg 2(LI^2|Z|\ell/\mathcal{F}\mu_r^2 r^3)$ and $(\omega_T \ell/c)^2 \gg 2(LI^2 Yx\ell/\mathcal{F}\mu_r^2|Z|r^3)$,

one gets $\beta \approx x/|Z|$, $\theta \approx \tan\theta = y/x$ and $\alpha \approx 0$. When the power supply is on and the charging phase is over, one gets in steady-state conditions $i = I$, hence,

$$d^2\theta/dt^2 + 2(\lambda - kd\psi/dt)d\theta/dt + [\omega_T^2 + 2(Xx/R^2)(c^2\ell/r^3)(LI^2/\mathcal{F}\mu_r^2|Z|)]\theta$$

$$\approx 2(Yx/R^2)(c^2\ell/r^3)(LI^2/\mathcal{F}\mu_r^2|Z|). \quad (12)$$

The initial condition of the pendulum is $\theta(0) = 0$,

thus,

$$\theta(t) \approx \theta_{max}[1 - e^{-\lambda t}\cos(\omega_0 t)], \quad (13)$$

$$\theta_{max} = (Y/X)[1 + \mu_r^2(\omega_T R/c)^2(r^3/\ell Xx)(\mathcal{F}|Z|/2LI^2)]^{-1}$$

$$\approx (Y/X)[1 + (\omega_S \omega_T R^2/c^2)^2(\mathcal{F}r^3\mu_r^2/2\ell XLI^2)^2]^{-1}, \quad (14)$$

$$\Omega(t) = d\theta/dt \approx \theta_{max}(\lambda^2 + \omega_0^2)^{1/2}e^{-\lambda t}\cos(\omega_0 t - \varphi). \quad (15)$$

Here, $\Omega(0) \approx \lambda\theta_{max}$, and also $\theta(t \to \infty) = \theta_{max}$ and $\Omega(t \to \infty) = 0$, where we have set $\tan\varphi = \omega_0/\lambda$ and $\omega_0 \approx (\omega_T^2 - \lambda^2)^{1/2}$; $d_{max} = 2\theta_{max} \times (MM' + M'M'')$.

When the power supply is off, after a duration of $5\tau$, the discharging phase is over. One gets in steady-state conditions,

$$\theta \approx \theta(t_0)e^{-\lambda(t-t_0)}\cos(\omega_0(t - t_0))$$

and

$$\Omega \approx -\theta(t_0)(\lambda^2 + \omega_0^2)^{1/2}e^{-\lambda(t-t_0)}\cos(\omega_0(t - t_0) - \varphi).$$

Note that $\Omega(t_0) \approx -\lambda\theta(t_0)$, which means that the pendulum starts to rotate in the opposite direction in accordance with the experiment and the prediction of [12] too.

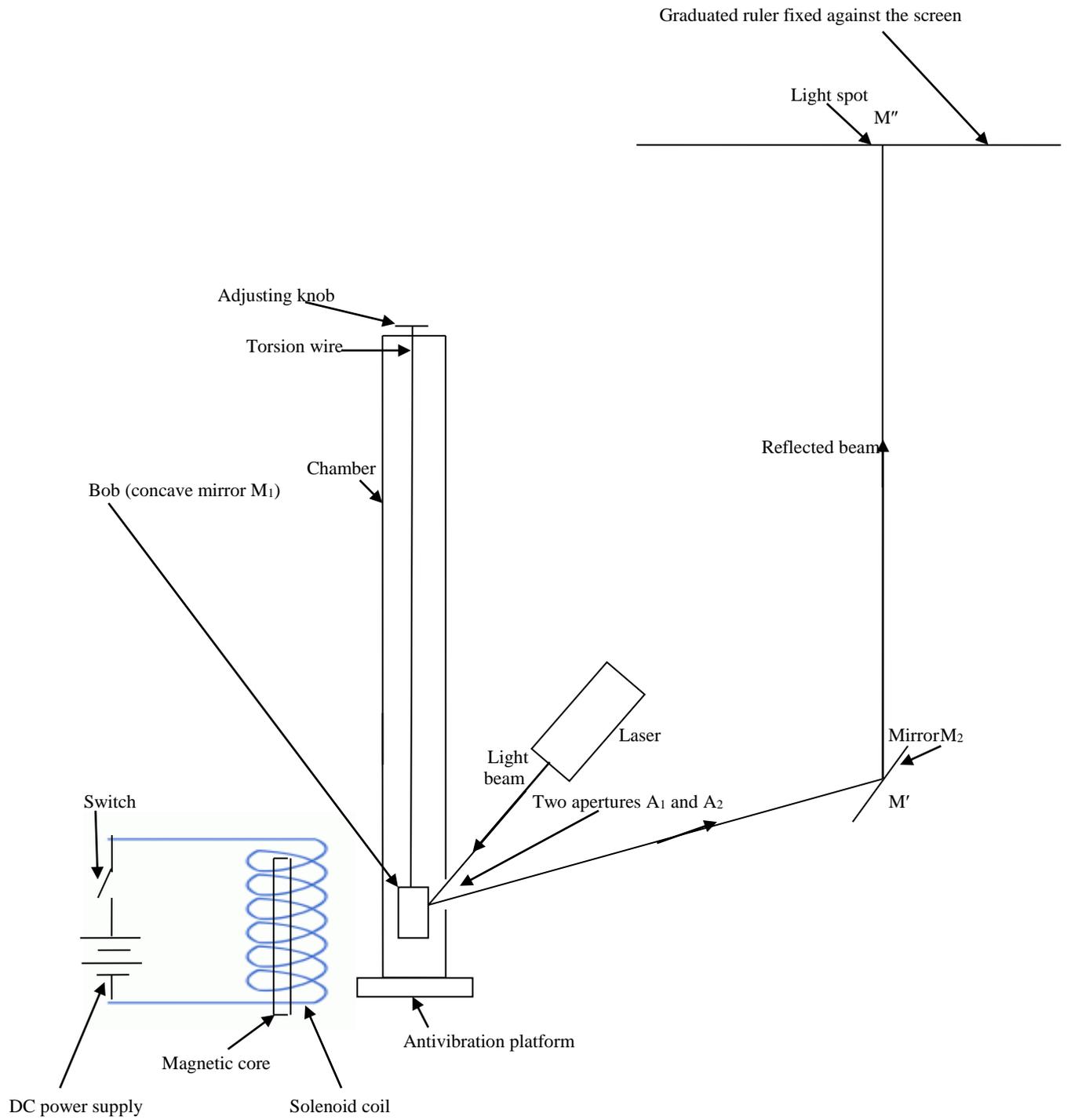

**FIG. 2.** Schematic of the experimental setup.

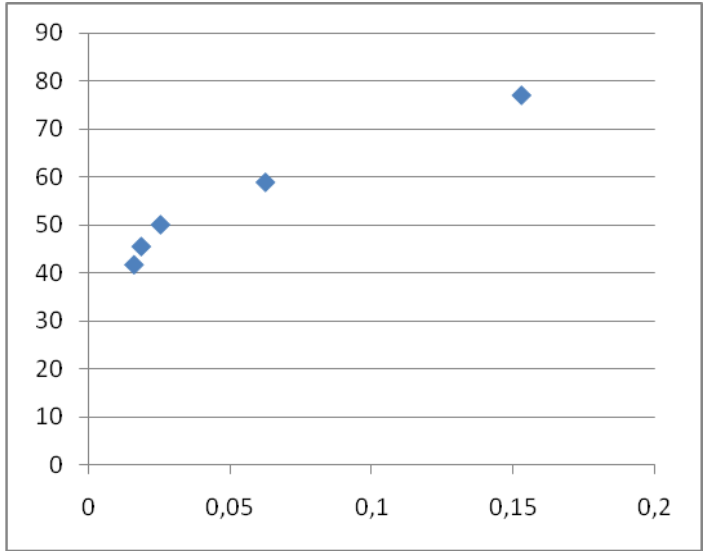

**Fig. 3.** Plot of $d_{max}(m)^{-1}$ versus $I(A)^{-4}$. One finds $d_{max}^{-1} = 209.9 I^{-4} + 45.07$ (see table 1), consistent with the theoretical prediction

$$\theta_{max}^{-1} = (r^2/4XY)(\omega_S \omega_T R^2/c^2)^2 (\mathcal{F} r^2 \mu_r^2 / l L)^2 I^{-4} + (X/Y),$$

for $X = -23\,cm$ and $Y = -0.3\,mm$ (a measurement gives $Y \approx -0.25 \pm 0.05\,mm$, see Fig. 9), with reduced chi-square $\chi_v^2 = 0.923$ on account of that $\delta d_{max} = 2\,mm$.

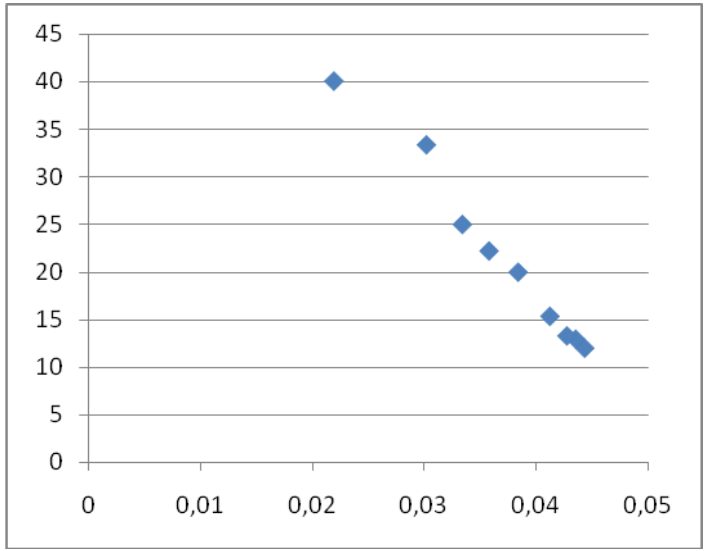

**Fig. 4.** Plot of $d_{max}(m)^{-1}$ versus $I(A)^{-4}$ during the decline of the current flowing through the coil (see table 2; the fit yields $\mu_r = 14.8 \times 10^3$, greater than previously, presumably because of hysteresis). One finds $d_{max}^{-1} = -1252\, I^{-4} + 67.2$, consistent with the theoretical prediction

$$\theta_{max}^{-1} = -(r^2/4XY)(\omega_S \omega_T R^2/c^2)^2 (\mathcal{F} r^2 \mu_r^2 / l L)^2 I^{-4} + (X/Y),$$

for $X = -23\,cm$ and $Y = -0.2\,mm$ (a measurement gives $Y \approx -0.25 \pm 0.05\,mm$), with reduced chi-square $\chi_v^2 = 0.810$ on account of that $\delta d_{max} = 2\,mm$.

**Table 1.** X = − 23 cm and Y = − (0.25 ± 0.05) mm

______________________________________
| I(A) | $d_{max}$(cm) |
|---|---|
| 1.6 | 1.3 |
| 2.0 | 1.7 |
| 2.5 | 2.0 |
| 2.7 | 2.2 |
| 2.8 | 2.4 |
______________________________________

**Table 2.** X = − 23 cm and Y = − (0.25 ± 0.05) mm

______________________________________
| I(A) | $d_{max}$(cm) |
|---|---|
| 2.6 | 2.5 |
| 2.4 | 3.0 |
| 2.34 | 4.0 |
| 2.3 | 4.5 |
| 2.26 | 5.0 |
| 2.22 | 6.5 |
| 2.20 | 7.5 |
| 2.19 | 7.7 |
| 2.18 | 8.3 |
______________________________________

IV. Discussion

One may wonder whether the magnetic field of the solenoid can generate some induced electric currents on the surface of the mirror thereby causing the observed torque on the pendulum. Now, from the Maxwell equation div **E** = $\rho/\varepsilon_0$, the current density **J** = $\sigma$**E** relation and the continuity equation div **J** + $\partial\rho/\partial t$ = 0, it follows

$\sigma\rho/\varepsilon_0 + \partial\rho/\partial t = 0$,

whose solution reads

$\rho(\mathbf{r},t) = \rho(\mathbf{r},0)e^{-(\sigma/\varepsilon_0)t}$.

Now, for aluminum (the mirror is made of glass and aluminum), one has $\sigma = 4\times10^7$ S·m$^{-1}$, thus

$$\rho(\mathbf{r},t) \approx 0, \mathbf{J} = \sigma\mathbf{E} = \sigma\mathbf{E}(\mathbf{r},0)e^{-(\sigma/\varepsilon_0)t} \approx 0$$

since $t \gg \sigma/\varepsilon_0 \approx 2.2\times10^{-19}$ s. As a consequence, the current $i$ through the surface of the aluminum mirror frame equals zero, and therefore the induced magnetic torque $\mathcal{M} = i\mathbf{S}\times\mathbf{B}$ equals zero too. Moreover, in as much as the axis of the solenoid and the normal to the mirror are parallel, the cross product $\mathbf{S}\times\mathbf{B}$ should be equal to zero. Thus the solenoid should not apply any torque to the bob of the pendulum in the case of a pure electrodynamics phenomenon where the photon is the sole fundamental boson brought into play.

Also, in order to check whether the observed motion of the pendulum might be due merely to an EM pulse generated by the power supply itself, the solenoid has been replaced by a pure resistor of resistance $R_{coil}$ (10Ω; 100 watts) so that exactly the same value of steady current, I, flows through the electric circuit. No torque applied to the bob was then observed.

A further important point consists in ensuring that no induced magnetic torques on the conducting component of the mirror may be responsible for the observed motion of the pendulum. In this respect, I have carried out the same experiment by using non-conducting media, namely glass and uncolored polymethyl methacrylate (PMMA) sheets. It turns out that the claimed effect is still present and obeys the same aforementioned law (see Figs. 5, 6, and 7). Let us point out that any other possible electromagnetic effect (e.g., residual ferromagnetic material response, paramagnetism or diamagnetism) should lead to a deviation that is proportional either to the current or its square. Now, it turns out that a linear or quadratic fit $d_{max}$ versus I is inappropriate, since it would yield a fictitious deviation even when the current is zero.

Another feature of the experiment described in this paper is that it reminds the Einstein-de Haas effect [13], though quite different in nature. Indeed, unlike the original Einstein-de Haas setup where a magnetic field is used to align the magnetic moments in a ferromagnetic material thereby causing its rotation in order to conserve angular momentum, in the experiment described in this paper, no magnetic material is involved whatsoever.

In addition, let us emphasize that we have done the same experiment described in section II but with an AC power supply in place of the DC one. In the case of a sine voltage, the deflection d changes sinusoidally with time, and a resonance is observed at 0.15 Hz by varying the frequency of the AC generator. Further results will be provided in a forthcoming paper.

**Table 3.** The coil axis parallel to the x axis;
equations of the coil axes : y = 0 and z = Z (X = – $M_0S$, Y ≈ 0 and Z = – OS).
The bob, of mass m = (35.5 ± 0.1) g, is a rectangular parallelepiped,
uncolored PMMA sheet (side length of 10 cm and thickness 2 mm)
attached to a wire of length $l$ = 17 cm; Δt denotes the measuring time

| U(V) | I(A)  | $d_{max}$(cm) | Δt(min) |
|------|-------|---------------|---------|
| 19   | 1.641 | 3.8           | 14      |
| 20   | 1.682 | 6.5           | 20      |
| 21   | 1.727 | 9.5           | 22      |
| 22   | 1.771 | 14.4          | 23      |
| 23   | 1.819 | 18.5          | 2       |
| 24   | 1.857 | 24.4          | 21      |
| 25   | 1.902 | 24.7          | 35      |
| 26   | 1.959 | 28.1          | 32      |

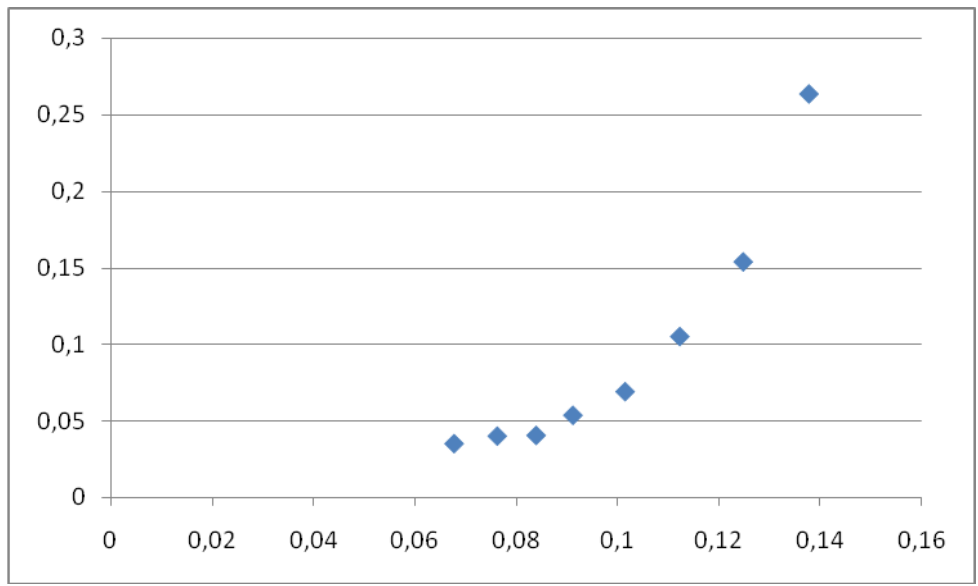

**Fig. 5.** Plot of $d_{max}$(m)$^{-1}$ versus I(A)$^{-4}$, one finds $d_{max}^{-1}$ = 298.7 $I^{-4}$ – 20.21 from the data in Table 3.

**Table 4.** The coil axis parallel to the y axis (equations of the coil axes : $x = -M_0S$ and $z = -OS$).
Such a configuration reduces significantly the magnetic flux through the bob.
However, the deviation is found much stronger than in the case of the mirror,
so that the distance from the bob to the screen is limited to $MM' = (3.22 \pm 0.01)$ m.
The bob, of mass $m = (35.5 \pm 0.1)$ g, is a rectangular parallelepiped, uncolored PMMA sheet
(side length of 10 cm and thickness 2 mm) attached to a wire of length $l = 17$ cm

______________________________

| U(V) | I(A) | $d_{max}$(cm) |
|---|---|---|
| 18 | 1.589 | 5.7 |
| 19 | 1.628 | 8.6 |
| 20 | 1.661 | 13 |
| 21 | 1.690 | 21.4 |
| 22 | 1.748 | 24.7 |
| 23 | 1.786 | 34.7 |
| 24 | 1.841 | 39.7 |
| 25 | 1.887 | 43.6 |
| 26 | 1.938 | 45.4 |
| 27 | 1.984 | 61.2 |
| 28 | 2.019 | 75.9 |

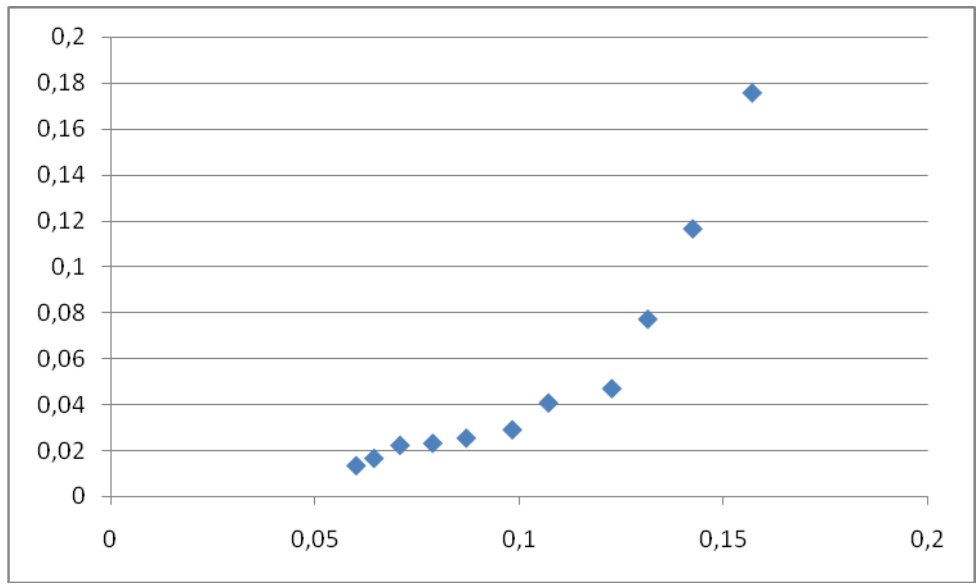

**Fig. 6.** Plot of $d_{max}$(m)$^{-1}$ versus $I$(A)$^{-4}$, one finds $d_{max}^{-1} = 138.2\ I^{-4} - 8.76$ from the data in Table 4.

**Table 5.** The coil axis parallel to the x axis and the coil above the chamber of the pendulum at about $SM_0 = 35$ cm height ($X \approx 0$, $Y \approx 0$, and $Z = OS$; equations of the coil axes : $y = 0$ and $z = OS$). Such a configuration allows no significant magnetic flux through the bob. The distance from the bob to the screen is $MM' = (3.22 \pm 0.01)$ m. The bob, of mass m = $(35.5 \pm 0.1)$ g, is a rectangular parallelepiped, uncolored PMMA sheet (side length of 10 cm and thickness 2 mm) attached to a wire of length $l = 17$ cm

______________________________________

| U(V) | I(A)  | $d_{max}$(cm) | $\Delta t$(min) |
|------|-------|---------------|------------------|
| 21   | 1.690 | 1             | 68               |
| 22   | 1.738 | 1.5           | 42               |
| 23   | 1.801 | 2             | 30               |
| 24   | 1.849 | 2.5           | 22               |
| 25   | 1.907 | 4             | 20               |
| 26   | 1.959 | 4.25          | 20               |
| 27   | 2.007 | 7.1           | 27               |
| 28   | 2.053 | 6.5           | 23               |
| 29   | 2.087 | 7.5           | 40               |
| 30   | 2.135 | 8             | 26               |
| 31   | 2.183 | 8.5           | 22               |

______________________________________

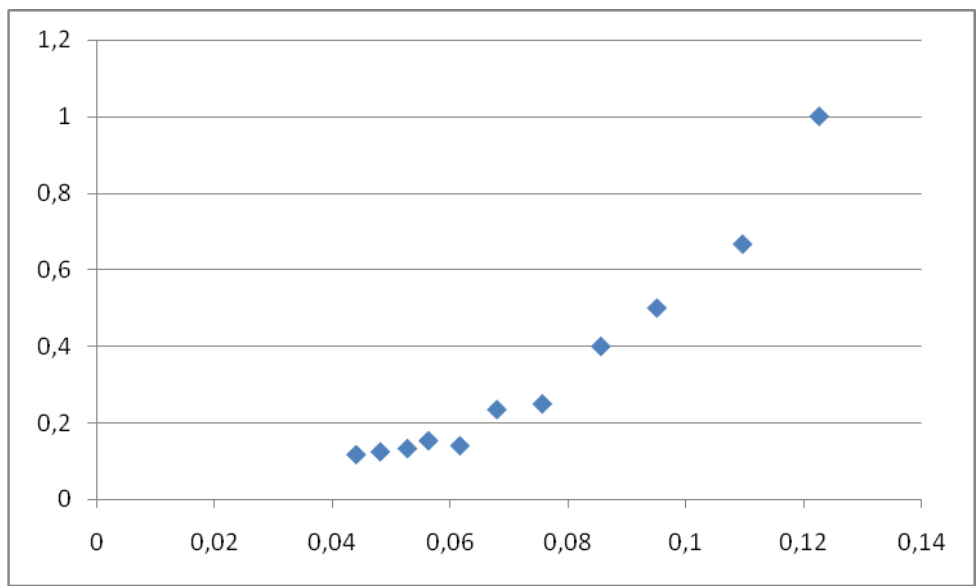

**Fig. 7.** Plot of $d_{max}(m)^{-1}$ versus $I(A)^{-4}$, one finds $d_{max}^{-1} = 1050\, I^{-4} - 44.33$ from the data in Table 5.

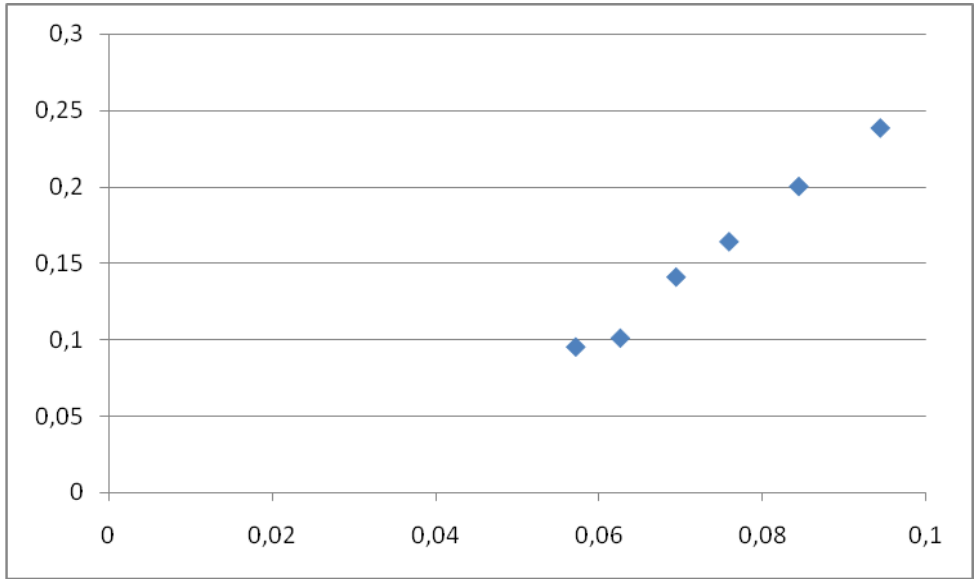

**Fig. 8.** Plot of $d_{max}(m)^{-1}$ versus $I(A)^{-4}$, with the axis of the coil being parallel to the y axis. The experiment was performed with a bob made of glass of mass m = (43.2 ± 0.1) g; Δt = 6 min to 23 min. One finds $d_{max}^{-1}$ = 401.7 $I^{-4}$ – 14.08, still consistent with the theoretical prediction. The current was varied from I = 1.756A to 2.090A.

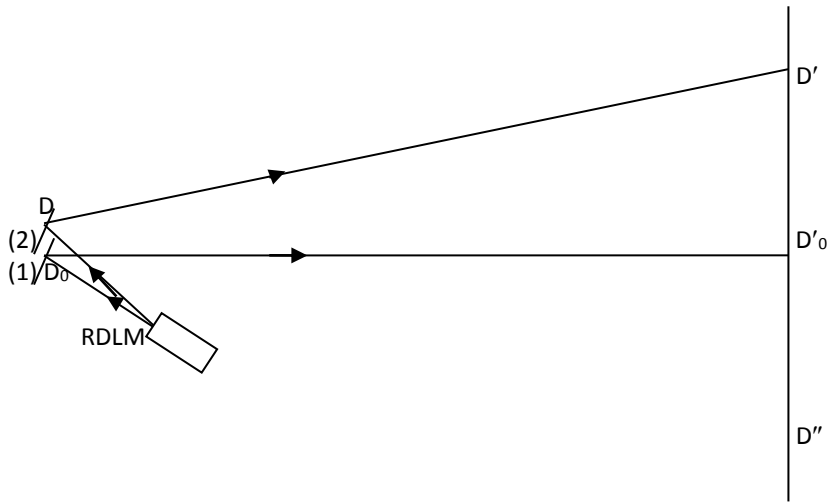

**Fig. 9.** Schematic of the experimental setup to measure the coordinate Y of the center of mass of the solenoid. A high-quality Red Dot Laser Module (RDLM) whose beam divergence amounts to 0.1mrad is used to measure Y. The laser beam from the RDLM is reflected on a small plane mirror mounted on the outer side of the solenoid. The laser beam is reflected by the mirror at point $D_0$ when the solenoid is in position (1) and point D when it is in position (2). In position (1), $S = S_0$, and the longitudinal axis of the solenoid is aligned to the normal of the pendulum bob. This is achieved with the help of the RDLM, after removing the magnetic core from the solenoid. The images of points $D_0$ and D on the screen are $D'_0$ and $D'$, respectively. Moreover, by using a compass, we draw a circle (C) with $D'_0$ as its center and a radius equal to $D'_0D'$. Thus, one obtains the other point of intersection, $D''$, between the circle (C) and the straight line ($D'_0D'$). The distances are $AD_0$ = 50cm and $D_0D'_0$ = 4.5m. The measurement yields $D'D''$ = (5 ± 1) mm and hence Y ≈ – (0.25 ± 0.05) mm, since $|Y| = S_0S = D_0D$, $D'_0D' = D'D''/2$ and $D_0D = D'_0D' \times AD_0/(LD_0 + D_0D'_0)$ according to Thales' theorem, where A stands for the aperture of the RDLM.
.

V. Conclusion

We have observed in the laboratory the expected gravitational effects of a quasistationary magnetic field, as predicted in the low velocity and weak field limit by a particular scalar-tensor theory of gravity, namely the KKψ theory. The experimental data are found in good agreement with the theoretical predictions. Many improvements to the present experimental setup and refinement of the measurements are still needed. In the meantime, on account of the simplicity of the experiment under consideration, hopefully some amongst experimentalists would find of interest to confirm or infirm the observations or conclusions put forward in this study.

Acknowledgment : I thank the anonymous referee for his valuable comments and suggestions, which helped in improving the paper.